\documentstyle[12pt,aaspp4]{article} 
\newcommand{\etal}{{\it{et al.}}~}

\newcommand{\eg}{{\it{e.g.}}}
\slugcomment{To be published in the Astronomical Journal}

\begin{document} 

\title{Diffuse Ionized Gas in Three Sculptor Group Galaxies}
\author{Charles G. Hoopes, Ren\'e A. M. Walterbos\altaffilmark{1}, and Bruce E. Greenawalt}
\affil{New Mexico State University, Department of Astronomy, Box
30001/Dept. 4500, \\ Las Cruces, New Mexico 88003 \\ choopes@NMSU.edu,
rwalterb@NMSU.edu, bgreenaw@NMSU.edu} 
\altaffiltext{1}{Visiting Astronomer, Cerro Tololo Inter-American
Observatory. CTIO is operated by AURA, Inc. under cooperative
agreement with the National Science Foundation.} 

\begin{abstract}

We present a study of the diffuse ionized gas (DIG) in three Sculptor
group galaxies: NGC 55, NGC 253, and NGC 300. The study is based on
narrow band imagery in H$\alpha+$[NII](6548+6583\AA) and [SII]
(6717+6731\AA). We find that DIG contributes 33 to 58\% of the total
H$\alpha$ luminosity in these galaxies, or 30 to 54\% after correcting
for scattered light. We find that NGC 300 has a higher fractional DIG
luminosity than the other galaxies in our sample, but it is not clear
whether this is a significant difference or an effect of the high
inclination of the other galaxies. The diffuse emission, averaged over
the optical extent of the disk, has a face-on emission measure of 5 to
10 pc cm$^{-6}$. The DIG is concentrated near HII regions, although
significant emission is seen at large distances from HII regions, up
to 0.5 to 1 kpc. The [SII]/(H$\alpha$ +[NII]) ratio is enhanced in the
DIG, typically around 0.3 to 0.5, compared to 0.2 for the HII regions
in these galaxies. These properties are similar to those measured for
the DIG in the Milky Way and in other nearby spirals. The line ratios,
large-scale distribution, and energy requirement suggest that
photoionization is the dominant ionization mechanism.

\end{abstract} 


\section{Introduction}

In recent years it has been recognized that a major component of the
Galactic interstellar medium is the diffuse ionized gas (DIG,
sometimes called the warm ionized medium or WIM). This gas has been
studied using several techniques, including pulsar dispersion
measurements (\eg $\space$ Reynolds 1991) and optical
line emission (\eg $\space$ Reynolds 1984, 1988). These observations
show the Galactic DIG to be widespread, warm (T $\sim$ 8000 K), and
diffuse (n$_e$ $\sim$ 0.2 cm$^{-3}$). It has a large vertical extent,
with a scale height of about 900 pc, fills 20\% of the Galactic
volume, and accounts for most of the mass of ionized gas (Reynolds
1991). The spectrum of the DIG differs from that of HII regions in
several respects. An example that is particularly relevant to this
study is the [SII] 6716+6731\AA/H$\alpha$ ratio, which is typically
about twice as high in the DIG as in HII regions (Reynolds
1985,1988). A large amount of energy, about 10$^{42}$ ergs s$^{-1}$,
is needed to keep this gas ionized, which can only be provided easily
by Lyman continuum photons from OB stars, and is just barely met by
supernova shocks (Kulkarni \& Heiles 1988).

Investigation of the DIG has recently been extended to external
galaxies. The vertical extent of the DIG has been studied in several
edge-on galaxies, such as NGC 891 (Rand, Kulkarni, \& Hester 1990;
Dettmar 1990; Keppel \etal 1991; Dettmar \& Schulz 1992), but few
face-on galaxies have been investigated. Detailed studies of
non-edge-on systems have been carried out only for M31 (Walterbos \&
Braun 1994, hereafter WB94, Walterbos \& Braun 1992), for NGC 247 and NGC
7793 (Ferguson \etal 1996), and for several irregular galaxies (Hunter \&
Gallagher 1990, 1992). An investigation of the DIG in M51, M81, and
several other nearby spirals is currently in progress (Greenawalt \&
Walterbos 1996). More limited analysis of the DIG has been done for
M33 (Hester \& Kulkarni 1990) and NGC 2403 (Sivan \etal 1990). In the
face-on systems that have been studied DIG appears to contribute 30 to
50\% of the total H$\alpha$ luminosity. The galaxies that have been
observed in [SII] show an enhanced [SII]/H$\alpha$ ratio similar to
that observed for the galactic DIG.

Although a general understanding of DIG characteristics is emerging,
the relatively small number of galaxies that have been studied makes
it difficult to draw universal conclusions, and several questions
remain unanswered. The connection of DIG with OB stars and HII
regions, and questions such as variation with galaxy type and
star-formation rate need to be investigated further to help constrain
the ionization mechanism.  We present a study of the DIG component of
three galaxies in the Sculptor group: NGC 55, NGC 253, and NGC
300. Our project is independent of the recent study by Ferguson \etal
(1996) of two other Sculptor spirals, NGC 7793 and NGC 247. The mean
distance to the Sculptor group is 2.5 Mpc, making these galaxies
ideally suited for this type of study. Table 1 presents some of the
properties of these galaxies.

In \S 2 of this paper we outline the observational procedure and data
reduction techniques, including tests for scattered
light contribution. The morphology of the DIG in the Sculptor galaxies
is described in \S 3. In \S 4, we calculate the luminosity of the DIG,
and in \S 5 we investigate the spectral characteristics. Finally, \S 6
contains a summary and discussion of the results.

\section{The Data}
\subsection {Observations and Data Reduction}

The data were obtained in September, 1990, with the 0.6 meter
Curtis/Schmidt telescope at CTIO. A Thompson 1024x1024 chip was used,
providing a field of view of 30.7x30.7 arcmin$^2$. The pixel size of
this chip is 1."84 when used with the Schmidt. Exposure times were 6
$\times$ 600 seconds in each filter for NGC 253 and NGC 55, and 6
$\times$ 540 seconds for NGC 300. We obtained images in each of three
filters: a 68\AA $\space$ wide H$\alpha$ filter, a 91\AA $\space$ wide
[SII] 6716+6731\AA $\space$ filter, and a 77\AA $\space$ wide
continuum filter centered on 6649\AA.  The H$\alpha$ filter also
contains the contribution from the [NII] 6548\AA $\space$ and 6584\AA
$\space$ lines. The spectro-photometric standards LTT 7987 and LTT
9293 were observed as well. NGC 253 and NGC 300 were observed under
photometric conditions, but NGC 55 was observed through cirrus. Table
2 is a log of the observations.

We reduced the CCD frames following standard procedures.  Dome flats
from various nights were co-added to create one ``super flat,'' which
we used to correct for gain variations across the chip.  We shifted
the images to a common grid, convolved to the worst seeing of the set,
and averaged using the {\it combine} task in IRAF. Cosmic ray events
were eliminated using the ``ccdclip'' option in {\it combine}. Bad
pixels were replaced with the median value of the neighboring
pixels. The continuum image was scaled to the line images (see \S 2.2)
and subtracted, producing net emission-line images. Remnants of bright
foreground stars were removed manually in IRAF and replaced by the
mean of the neighboring pixels. The NGC 253 and NGC 300 images were
calibrated using the observed spectrophotometric standards (Hamuy et
al. 1992). NGC 55 was taken under non-photometric conditions. We
derived a calibration for NGC 55 using the published R-band magnitudes
(Alcaino \& Liller 1984) of stars in the field.

\subsection {Uncertainties}

The rms noise in the images, measured on the background, is given in
table 2. One must also be concerned about flat fielding errors when
working with continuum subtracted images, which can produce variation
over large regions in the net line images. By comparing individual
flat-field images, we estimate that a 1\% flat fielding uncertainty
may exist over the region of the chip containing the galaxies. This
error is in both the line and continuum images, so when the continuum
is subtracted the uncertainty may be compounded. For NGC 253 a 1\%
error in the flat fielding could produce a systematic uncertainty in
the worst case of about 4.6 pc cm$^{-6}$ when comparing one side of
the disk to another. For NGC 300 the uncertainty is 3.9 pc
cm$^{-6}$, and 4.6 pc cm$^{-6}$ for NGC 55.

During the observations we experienced problems with horizontal
banding structure in the bias level of the CCD. Usually the bands were
low-level, on the order of a few counts, but some images contained
jumps of tens or hundreds of counts.  The NGC 253 and NGC 300 images
were corrected satisfactorily with overscan subtraction, but the
problem was worse on the night that NGC 55 was observed. We manually
removed the large bands from the image, and corrected for the smaller
jumps using a higher order fit to the overscan region of the chip.
Our efforts substantially increased the quality of the images, but
some banding is still evident in the final images. This is the origin
of the diagonal pattern in the background of the NGC 55 image (figure
1). The remaining structure is very low level (less than 3 pc
cm$^{-6}$ in emission measure). It is obvious in the figure because
the image is displayed so that the lowest levels can be seen to
emphasize the diffuse emission.

A critical procedure in the data reduction is the continuum
subtraction. To determine an initial scale factor with which to
multiply the continuum image to scale it to the line images, we
measured the fluxes of about 20 stars in the frame of each galaxy in
each filter, and calculated the multiplicative factor needed to make
the mean of the stellar fluxes in different filters equal. However,
the foreground stars do not necessarily reflect the continuum spectrum
of the galaxies. We also determined scaling factors by comparing
regions in each galaxy which appeared to be free of line
emission. This only led to adjustments for NGC 300, where the net
H$\alpha+$[NII] image produced using the scale factor from foreground
stars became negative in certain regions, while the net [SII] image
still appeared to contain significant continuum emission. Adjusting
the scale factor of the continuum image by $-2$\% for H$\alpha+$[NII]
image, and by $+6$\% for the [SII] (as indicated by the line-free
galaxy regions) seemed to produce more plausible results. The scale
factor derived from the foreground star would have produced
unrealistically high [SII]/H$\alpha$ ratios. In view of this, we
estimate that the scale factor used in subtracting the continuum may
be uncertain at the $\pm 3$\% level and wherever appropriate we quote
the corresponding uncertainties.

The southwest side of NGC 253 (Figure 2) appears weaker in H$\alpha$
than the rest of the galaxy. This is the side that is red-shifted due
to rotation, with a maximum redshift of about 5\AA. Combined with the
5\AA $\space$ systemic redshift of NGC 253, the H$\alpha$ line is then
shifted up to 10\AA $\space$ with respect to the center of our
filter. In addition, the F3.5 beam of the Schmidt telescope causes the
filter passband to be blue shifted about 10\AA. While H$\alpha$ is
still well within the filter, [NII] at 6584\AA $\space$ is shifted to
about the 50\% transmission level. The 5\AA $\space$ rotational
redshift of NGC 253 pushes the line further toward the edge of the
passband for the southwest side of the galaxy. This may be the cause
of the difference in appearance of the two sides of the galaxy, and
may affect our determination of the total H$\alpha$
luminosity. Kennicutt (1992) finds an average [NII]/H$\alpha$ ratio
for spiral disks of 0.5. Assuming this is also true for NGC 253, a
loss of half of the flux from [NII] 6584 would reduce the observed
H$\alpha$ +[NII] flux by 17\% on that side of the disk. It should not
affect the other two galaxies, whose systemic velocities produce
redshifts of only 3\AA $\space$, and whose rotational velocities are
much less than that of NGC 253 (Tully 1988).

\subsection {Scattered Light}

It is crucial to our study of the DIG that we understand the
contribution of scattered light.  Light scattered by the optics in the
telescope and camera produces wings in the point spread function,
which might be interpreted as low-level emission surrounding bright
objects like stars or HII regions. Emission by HII regions may also be
scattered by dust in the galaxy, again producing diffuse emission
around HII regions. We tested for these effects following WB94.

We determined the point spread function (PSF) for two bright stars on
images in both filters, to see how much of the stellar flux was
scattered into the extended tail. The resulting stellar profiles
extend over a range of more than 10 mag arcsec$^{-2}$. In these
profiles, 95\% of the measured flux from the star is contained within
9 pixels (16'') in the H$\alpha$ image, and within 13 pixels (24'') in
the [SII] image. Light scattered in the telescope would be
concentrated within 16'' of HII regions (in H$\alpha$), and not more
than 5\% of the total H$\alpha$ flux would come from beyond this
radius around HII regions. However, we find that the actual H$\alpha$
contribution of the DIG is substantially larger than this. To test
this quantitatively, we made an image of the PSF using unsaturated
foreground stars in the images. We convolved the PSF with an image
containing only the HII regions, with the DIG masked out, to simulate
the scattering of light due to HII regions. We then masked out the HII
regions from the convolved image using the method discussed in \S4.2,
which left only the light scattered out of the HII regions, and
compared the scattered light to the diffuse emission in the original
images. As an example, in NGC 253 the contribution of the DIG to the
total H$\alpha$+[NII] emission was reduced from 40\% to 33\% if the
scattered light is removed, indicating that 18\% of the diffuse
emission is in fact scattered light. The DIG contribution to the total
[SII] luminosity drops from 54\% to 50\%, a reduction of 7\%. In NGC
300, the diffuse H$\alpha$+[NII] emission was reduced by 8\% when this
correction was made, and the diffuse [SII] emission dropped by 7\%. In
NGC 55, the H$\alpha$+[NII] correction was 8\%, and the [SII]
correction was 6\%. We conclude that up to 20\% of the DIG flux in
H$\alpha$ may in fact be due to scattered light, but only 10\% in
[SII]. The fraction is less in [SII] even though the PSF appears more
extended, because the [SII] emission from HII regions is much
fainter. We have applied this correction to our measured DIG
fractions, but we will also give our results without the correction
for comparison with previous studies.

Light scattering by dust in the galaxy as the dominant source of the
diffuse emission is unlikely because the [SII]/H$\alpha$ ratio
observed in the DIG is quite different from that observed in HII
regions. Additionally, light scattered by dust would usually be
concentrated around HII regions, but the DIG emission can occur in
distinct patches or filaments far from HII regions.

\section{Morphology of the DIG}

The continuum subtracted H$\alpha$ images of NGC 55, NGC 253,
and NGC 300 are shown in figures 1-3. The images are calibrated
in emission measure and displayed logarithmically to show the faint
emission of the DIG. These images show that diffuse ionized gas is
clearly present in these galaxies. 

As noted earlier, NGC 253 is a prototypical nuclear starburst. The
likely heavily obscured H$\alpha$ luminosity of the nucleus alone is
about 6 $\times$ 10$^{38}$ ergs s$^{-1}$ (measured in an 8.1''
aperture, which corresponds to a diameter of 63 pc). To the south and
east of the nucleus is the outflow cone, powered by stellar winds and
supernovae in the starburst nucleus (Schulz \& Wegner 1992; Heckman
\etal 1990 and references). There are many bright HII regions,
especially in the central 7 kpc of the galaxy, along the two
well-defined spiral arms. Emission from diffuse ionized gas is
distributed between the HII regions along the arms, but is strongest
close to HII regions. In the outer part of the galaxy the DIG is
usually associated with a specific HII region, and in the inner parts
where there are many HII regions the DIG is spread throughout the arms
and even into the inter-arm region. At the limited resolution of the
data, the DIG structure is primarily diffuse, although some
filamentary structure can be seen, especially on the northwest edge of
the galaxy.

NGC 300 is quiescent in comparison to NGC 253, as shown by the
relatively small number of bright HII regions and lower H$\alpha$
luminosity (Table 3). Diffuse emission is again primarily associated
with HII regions. Here it can be seen that the morphology is a
combination of diffuse and filamentary structures. The filamentary
structure of the DIG in NGC 300 is striking. Loops or shells, possibly
remnants of expanding shells which were driven by the stellar winds
and supernovae of OB associations, dominate the morphology. Some of
these loops are associated with star-forming regions, while others are
completely isolated from any bright HII regions. Both complete and
partial shells are visible.

The continuum subtracted H$\alpha$ image of NGC 55 also shows diffuse
and filamentary emission from DIG.  Again, the trend is for diffuse   
emission to be concentrated near star-forming regions, and to decrease
in intensity as one moves away from HII regions. DIG is most evident in
the central 4 kpc of the galaxy, where the largest HII regions are
also seen. This image also shows diffuse emission extending vertically
from the plane of the galaxy. On both sides the emission can be traced
vertically more than 1.5 kpc from the plane. 

The central region and halo of this galaxy shows prominent filamentary
structure in the DIG. A large emission-line loop north and east of the
nuclear bulge has been observed previously in [OIII] (Graham \& Lawrie
1982). This loop is present in our H$\alpha$ image but not in
[SII]. The [SII] image has similar noise than the H$\alpha$ image, but
the expected signal will be weaker so the loop may be below our limit
of detection. The upper limit to the [SII]/H$\alpha+$[NII] line ratio
is 0.25. We do detect another loop on the south side of the galaxy in
both H$\alpha$ and [SII]. This loop is smaller but more obvious than
the north loop in our H$\alpha$ image. In addition, more loop- and
shell-like structures are seen in H$\alpha$ east of the southern loop,
and a large filament extends south of the galaxy on the west side of
this southern loop in both filters. This filament can be traced for
1.5 kpc in H$\alpha$. The loops and filaments may be related to
stellar wind and supernovae driven shells, similar to those in NGC
300, and indicate a connection of the disk ISM to the halo gas.
	
In summary, all three galaxies show emission from diffuse ionized gas. The
distribution is both filamentary, as in the loops and shells, and a
diffuse layer in the inner disks. Qualitatively, there is a
correlation between the distribution of DIG and that of HII regions,
implying a connection with star formation. In the following sections we
will explore this connection in more detail.

\section{Luminosity of the DIG}

\subsection{Total H$\alpha$ Luminosity}

The H$\alpha$+[NII] luminosities of each galaxy were calculated by
integrating pixel values over the image. We removed foreground stars
from the image and then set the background to zero by subtracting a
plane fit to the border of the image. The derived luminosities
are given in table 3. These galaxies are located well out of the plane
of the Milky Way ($b$ $\le -75^\circ$), so the correction for
foreground extinction is small (Burstein \& Heiles 1984), much less
than our observational uncertainty. We also did not correct for
internal extinction, which could have a significant effect in NGC 253
and NGC 55, but is difficult to quantify.

The uncertainties in table 2 were derived by varying the continuum
subtraction by $\pm$ 3\%.  The luminosity for the nucleus of NGC 253
is 6.3$\times$ $10^{38}$ ergs s$^{-1}$, 13\% lower than that measured
by Keel (1984), including a rough correction for the shifting of the
[NII] 6584\AA $\space$ line in the filter. This is the only
independent determination of the H$\alpha$ flux from NGC 253 that we
found in the literature.

We calculated the star-formation rate (SFR) from the total H$\alpha$
luminosity (Kennicutt 1983), giving about 0.2 $M_{\odot}yr^{-1}$ for
NGC 55, 0.4 $M_{\odot}yr^{-1}$ for NGC 253, and 0.1 $M_{\odot}yr^{-1}$
in NGC 300. These values are very uncertain, especially in NGC 253 and
NGC 55, due to internal extinction. Although the SFR for NGC 253
derived from the H$\alpha$ luminosity is only a factor of four greater
than that of NGC 300, the FIR luminosity (Rice \etal 1988) rescaled to
our adopted distance gives a L$_{FIR}$/L$_{H\alpha+[NII]}$ ratio of
178 for NGC 253, but only 15 for NGC 55 and 12 for NGC 300. This
discrepancy probably indicates that most of the H$\alpha$ emission in
NGC 253 is obscured.  High resolution FIR imaging of NGC 253 (Smith \&
Harvey 1996) show that $\geq$ 70\% of the FIR luminosity arises from
the starburst nucleus. The H$\alpha$ emission from the nucleus is
likely heavily obscured by dust, and the disk itself is known to be
extremely dusty (Sofue \etal 1994, Sandage 1961), so it is not
surprising that the H$\alpha+$[NII] luminosity is low compared to the
FIR luminosity.  The SFR of NGC 253 may be a factor of ten or more
higher than that indicated by the H$\alpha$ luminosity. It is clear
that these galaxies have widely varying star formation properties.

\subsection{Isolating the Diffuse Emission on the Images}

Perhaps the most reliable method to estimate the H$\alpha$ flux
contributed by the DIG in a galaxy is to subtract from the total
H$\alpha$ flux the contribution from discrete HII regions and other
emission line sources (e.g., WB94). However, this method requires
identification and individual flux measurements, using aperture
photometry with correction for local background fluxes, for several
hundred to as much as a thousand objects. Given the difference in the
surface brightness of the DIG compared with discrete emission line
nebulae, it should be possible to get the desired number in a more
straightforward way by masking out discrete sources before summing up
the flux. A possible complication, however, which is especially severe
in strongly inclined galaxies, is that the discrete source population
is superposed on a potentially varying DIG background. This implies
that a simple cut at a single surface brightness level (e.g., Ferguson
\etal 1996) to isolate DIG from HII regions is not a suitable method
in all cases. As an example, in Fig. 4a, we show the image of NGC 253
cut an isophotal level of 80 pc cm$^{-6}$, the level chosen by
Ferguson \etal for NGC 247 and NGC 7793. The entire inner disk of the
galaxy has been clipped, whereas inspection of Fig. 2 shows that DIG
is clearly present in this region. Cutting at a higher intensity level
results in leaving in too many HII regions in the outer disk.

A related problem is that the DIG intensities are likely to increase
with inclination, simply due to projection effects. Cutting all
galaxies in the sample at a constant isophotal level would therefore
lead to DIG fractions that will have potentially large systematic
differences that are mostly the result of the method used to measure
the DIG. To circumvent this, we developed a slightly more
sophisticated masking technique to isolate the DIG.

The H$\alpha$ image of each galaxy was median filtered with a large
box (90 $\times$ 90 arcseconds) to remove the small structure in the
image and leave only the large scale variations. We experimented with
different sized boxes on the Sculptor galaxies and on the M31 images
of WB94. These images are more sensitive and have better spatial
resolution than our Sculptor group images. In WB94 it was therefore
possible to calculate an accurate DIG luminosity using the source
subtraction technique. We chose the box size that gave diffuse
fractions for the M31 fields that were on average consistent with
those found in WB94 and applied a box with the corresponding linear
size to the Sculptor galaxies. The median filtered image was then
subtracted from the original to remove the large scale emission across
the galaxy, leaving only the small scale structure. The resulting
image was slightly smoothed to reduce the noise. A mask was defined on
this image by making a cut at a constant surface brightness level of
50 pc cm$^{-6}$. Pixels below the limit were replaced with a
value of one, and those above were replaced with zero. This mask was
then multiplied with the original H$\alpha$ image, resulting in an
image of the DIG component of each galaxy, with pixels in HII regions
repaced with zero flux. The pixel values in the masked image were
added to get the total emission from the DIG. This was compared to the
total H$\alpha$ luminosity to get the contribution of DIG emission to
the total H$\alpha$ luminosity of each galaxy.

The DIG flux measured from the masked image is an underestimate,
because we are neglecting diffuse emission that may be superimposed on
the masked HII regions. To correct for this, we estimated the mean
intensity of the DIG emission in the galaxy, and multiplied this with
the number of masked pixels. This may still be an underestimate,
because the DIG emission near HII regions may well be higher than the
mean. We also corrected for the scattered light contribution to the
DIG using the method discussed in \S2.3.

We found that this method produced diffuse fractions for the inclined
galaxies that are relatively insensitive to the cutoff level used on
the median-subtracted image. The diffuse fraction for NGC 300 is still
somewhat sensitive to the cut level. Our cut at 50 pc cm$^{-6}$
results in a relatively high diffuse fraction of 53$\pm5$\%. If
the cut is made at 30 pc cm$^{-6}$, the diffuse fraction is 45$\pm5$\%
for NGC 300. A masked image of NGC 253 made with this method is shown
in figure 4b for illustration. This mask is much better at
differentiating between the DIG and HII regions in both the faint
outer disk and the bright inner disk. Comparison with figure 4a shows
that this method works better than a cut at constant surface
brightness, especially for inclined galaxies such as NGC 253.

We also investigated whether a large fraction of the DIG luminosity
might be contributed by faint unresolved HII regions. To test for
this, we estimated the faintest HII regions that could be
distinguished from the background and hence masked out of the
images. This luminosity limit is about 10$^{36}$ erg s$^{-1}$ for all
three galaxies. Using the HII region luminosity function derived by
Kennicutt \etal (1989), we estimated the contribution of HII regions
below this limit to the DIG flux. The contribution is less than 5\% in
all three cases.

\subsection{The Diffuse Fractions and Mean Emission Measure}

The measured diffuse fractions are given in table 3. The range was
found by varying the continuum subtraction by $\pm$3\%. The diffuse
fraction measured in NGC 55 and NGC 253 could still be a lower limit,
due to the difficulty in isolating DIG from HII regions in inclined
galaxies. It is therefore not clear that the somewhat lower fractions
in these two systems indicate a real difference with NGC 300. With
this in mind, the fraction of H$\alpha$ luminosity arising in the DIG
appears to be remarkably constant, especially given the varied nature
of these galaxies. The diffuse fractions measured here are similar to
those measured in other nearby galaxies, such as M31 (WB94), M51 and
M81 (Greenawalt \& Walterbos 1996), and NGC 247 and NGC 7793, two
other Sculptor group galaxies (Ferguson, \etal 1996).

While we believe that the fractional DIG fluxes measured this way are
accurate, we have also derived numbers using a simple cut at a
single isophotal level. This showed that the difference in DIG
fraction between the galaxies in our sample and that of Ferguson \etal
(1996) can be easily attributed to the measurement technique. A
disadvantage of the median filter method is that we lose the ability
to determine how the DIG flux increases if cuts are made at
successively higher intensities. A reason for doing this would be to
compare the estimates for the Galaxy, which are based on solar
neighborhood data, with external systems. In particular, all external
systems looked at so far indicate {\it higher} fractional Lyman
continuum requirements than determined for the Galaxy. However, we
believe that this is likely due to the different level of the DIG
present in the solar neighborhood compared to the average galaxy
disk. In particular, in figure 5 we show how the fractional DIG flux
increases as we increase the isophotal cutoff level to mask out the
DIG. Our conclusion is that the relatively low estimate (10 to 15\%,
Reynolds 1991b) of the fraction of Lyman continuum photons required to
ionize the DIG in the Galaxy, compared to the results obtained for
external systems, is most likely due to the presence of the Sun in a
region of rather low DIG surface brightness, with an EM of about 5 to
6 pc cm$^{-6}$, while the average values in spiral arms and even
across entire disks, are substantially higher (see table 3). If we
restrict the DIG in the Sculptor galaxies to face-on EM levels below 6
pc cm$^{-6}$ we also find 10 to 15\%, rather than 30 to 50\%.

The average emission measure in the DIG was determined within the 25th
blue magnitude isophote on the inclination corrected disks of NGC 253
and NGC 300. For NGC 253, the average EM in the DIG is $6.1-9.7$ pc
cm$^{-6}$. The range in values is the results of varying the continuum
subtraction by $\pm$3\%. The average emission measure for the nominal
continuum subtraction is 7.9 pc cm$^{-6}$. For NGC 300 the range is
$4.6-7.9$ pc cm$^{-6}$, and the nominal value is 6.7 pc cm$^{-6}$. A
$cos$ $i$ correction was applied to NGC 300 and NGC 253, so these
numbers indicate face-on surface brightness. This is probably an
over-correction due to dust extinction in the disk, the vertical
extent of the disk, and the DIG concentration in spiral arms. For NGC
55 we measured the average emission measure in a 20.24' $\times$ 4.45'
box centered on the galaxy in the masked image, parallel to the major
axis. We then deprojected the observed average emission measure using
a simple approximation of a uniformly-filled cylinder. The
face-on average emission measure using this simple model is
$8.5 - 11$ pc cm$^{-6}$. These values are lower limits to the true
average surface brightness of the DIG, because the area considered is
not completely filled with DIG. The average emission measure is
certainly higher in the spiral arms. In the solar neighborhood the
average emission measure of the DIG through the disk is 5.6 pc
cm$^{-6}$ (Reynolds 1984). In M31, WB94 find an average EM of $6-15$
pc cm$^{-6}$. Our values are consistent with these measurements.

Figure 6 shows the radial distribution of HII region and DIG emission
for NGC 253 and NGC 300, corrected for inclination. The difference in
the H$\alpha$ luminosity of these galaxies is apparent, especially in
the nuclear region. The radial profiles show in a general sense the
correlation of the DIG distribution with the HII region
distribution. In the outer disks of both galaxies, the DIG appears to
contribute a larger portion of the total H$\alpha$ luminosity. Note
that this effect occurs in the faintest part of the galaxy, where the
emission measure is less than 10 pc cm$^{-6}$. This part of the galaxy
is very sensitive to continuum subtraction errors, background errors,
and flat-fielding errors, as can be seen by the error bars of figure
6. The loss of sensitivity in this region precludes us from claiming
that the diffuse fraction varies with radius. Ferguson \etal (1996)
have found an increase in the diffuse fraction with radius in one of
the galaxies in their sample, NGC 7793. The inner and intermediate
disks of NGC 300 and NGC 253, where the uncertainty is less, show a
DIG contribution that is fairly constant.

The contribution of DIG to the total H$\alpha$ luminosity of
these galaxies seems to be a constant fraction of the HII region
luminosity, both within and among different galaxies. This and the
noted correlation with star-forming regions implies a close
correlation of massive stars with the DIG.

\section{Spectral Properties of the DIG}

In this section we investigate the [SII]/(H$\alpha$+[NII]) ratio of
the DIG, which may help in constraining the ionization mechanism. We
applied the mask (see \S 4) to both the H$\alpha$ and [SII] images to
compute the diffuse fractions in each line. The ratio of the resulting
fluxes is shown in table 3. The same was done for HII region
emission. These values were calculated after the correction for
scattered light. The ratio in HII regions is about 0.2, while that measured
in the DIG is about 0.35. The error bars indicate the variation due to
continuum subtraction, and mostly reflect the sensitivity of the [SII]
emission to the continuum scale factor. The [SII] luminosity of NGC
253 is extremely dependent on the continuum subtraction, varying by
$\pm$55\% when the continuum is varied by $\pm$3\%. Varying the
continuum scale factor changed the [SII] luminosity of NGC 55 and NGC
300 by $\pm$19\% and $\pm$35\%, respectively, so this effect is less
serious for these galaxies. Our measurements also contain some [NII]
emission. If the [NII]/H$\alpha$ ratio is also higher in the DIG, as
some studies (Keppel \etal 1991; Dettmar \& Schulz 1992) have
indicated for NGC 891, then the actual difference in the
[SII]/H$\alpha$ ratios is higher than our measurements indicate. For
example, if the [NII]/H$\alpha$ ratio in the DIG were 0.6 (the value
obtained by Dettmar \& Schulz (1992) for a height of 0.5 kpc above the
plane), compared to 0.3 in HII regions, the corrected [SII]/H$\alpha$
ratio for HII regions would be 0.26, but the ratio in the DIG would be
0.56. Dettmar \& Schulz (1992) saw [NII]/H$\alpha$ ratios as high as
1.1 at z=1 kpc above the plane, further enhancing the difference.
However, no increase in the [NII]/H$\alpha$ has been seen in the
Galactic studies (Reynolds 1990).

To further investigate the line ratio in the DIG, we plot in figure 7
the intensity ratios of regions in the diffuse gas and in HII regions
as a function of H$\alpha$ intensity for the Sculptor galaxies. Due to
the near face-on orientation of NGC 300, emission measures up to 40 pc
cm$^{-6}$ in NGC 300 should be compared with emission measures up to
100 pc cm$^{-6}$ in the inclined galaxies. The regions were measured
in 9''$\times$9'' boxes. The HII regions (figure 7 a-c) consistently
show [SII]/(H$\alpha$+[NII]) ratios around 0.2, with small 1$\sigma$
error bars. The diffuse gas (figure 7 d-f) shows a higher ratio,
typically around 0.3 to 0.5. The error bars in each plot are a
combination of photon noise and continuum subtraction variation. The
large error bars for NGC 253 again reflect the sensitivity of the
[SII] luminosity to the amount of continuum subtracted in this
galaxy. The ratio obtained for HII regions is contaminated by
foreground DIG. This tends to overestimate the observed ratio,
especially in faint HII regions.

Investigation of the ratios in individual boxes shows the change in
the [SII]/(H$\alpha$+[NII]) ratio between HII regions and DIG. The
high ratio is present in the diffuse layer around HII regions, in the
inter-arm emission of the inner disk of NGC 253, and in the discrete
structures such as the loops and shells in NGC 300 and NGC 55. The DIG
projected within the shells shows the same spectral signature. In NGC
55 we note that the enhanced ratio exists in the diffuse emission
above the plane of the galaxy. However, our images are not sensitive
enough to tell if the ratio changes with height above the plane. There
is a suggestion that the [SII]/(H$\alpha$+[NII]) ratio increases with
distance from HII regions, but this result is not conclusive given the
signal to noise ratio of the data and the systematic errors
present. However, spectroscopy of the DIG in M31 has shown this effect
(Greenawalt, Walterbos \& Braun 1996).

For the Galaxy the [SII]6716\AA/H$\alpha$ ratio has been observed to
vary from 0.07 to 0.18 in HII regions, with the higher ratio
corresponding to the most dilute HII regions (Reynolds 1988). The
observed ratios for DIG (or Warm Ionized Medium) in the Galaxy are 0.3
to 0.5 (Reynolds 1985) corresponding to a
[SII]6716+6731\AA/H$\alpha+$[NII]6548+6583\AA ratio of 0.35 to
0.57. The trend for higher [SII]/H$\alpha$ ratios has also been seen
in the DIG of other galaxies such as M31 (WB94) and M51 and M81
(Greenawalt \& Walterbos 1996). Our results for the Sculptor galaxies
are consistent with those reported previously. The high
[SII]/H$\alpha$ ratio can occur in a shock ionized environment,
suggesting that shocks from supernovae may be an important ionization
source. However, the ratio is critically dependent on the shock
velocity (Mathis 1986). Overall, the [SII]/H$\alpha$ ratio seems to be
rather constant in the DIG, which would seem to indicate an
unrealistically constant shock speed (Mathis 1986). In addition, the
energy requirement of the DIG in several galaxies is higher than can
be provided by supernovae (WB94; Rand \etal 1990). Photoionization in
a diffuse medium (Mathis 1986) can also produce the observed ratio,
with perhaps a contribution from photoelectric heating by dust grains
(Reynolds \& Cox 1992; Sivan \etal 1986). If photoionization is the
mechanism responsible for the DIG, the structure of the ISM, and
particularly of neutral Hydrogen, must be better understood to make it
clear how ionizing photons can travel the large distances required to
reach the DIG without being absorbed.

\section{Summary and Discussion}

The data presented in the preceding sections show the presence of DIG
in NGC 55, NGC 253, and NGC 300. The contribution of DIG emission to
the total H$\alpha$ luminosity is observed to be 33 to 58\% in all
three galaxies, uncorrected for extinction. After correcting for
scattered light, the diffuse fraction is between 30 and 54\%. The
average emission measure of the two non-edge-on galaxies is 6.2 pc
cm$^{-6}$ for NGC 300 and 6.6 pc cm$^{-6}$ for NGC 253. The diffuse
emission is strongest near HII regions and in the inner disks of these
galaxies, and is observed both as a diffuse layer and with filamentary
structure in the form of shells and loops. The global distribution of
DIG in the disks seems to be correlated with the HII region
distribution, implying a connection with star formation. The
[SII]/(H$\alpha$ +[NII]) ratio is enhanced in the DIG compared to HII
regions, with ratios of 0.3-0.5, compared to around 0.2 for HII
regions. All of the derived characteristics compare well with the
values seen in the Milky Way and in other spirals galaxies.

One of the surprising discoveries about the DIG in spiral galaxies is
the constancy of the fraction of the total H$\alpha$ luminosity of a
galaxy that it contributes. WB94 measured 40\% for M31, and all
results obtained since have come up with a similar number: for the
galaxies studied here, we find fractions between 30 and 54\%, for M51
and M81 we find 30 to 40\% (Greenawalt \& Walterbos 1996), while
Ferguson \etal (1996) find 53\% for NGC 247 and 41\% for NGC 7793. If
it were not for the difference in [SII] over H$\alpha$ intensity ratio
in the DIG compared to HII regions, we might in fact question whether
we are mostly detecting H$\alpha$ photons that were scattered into the
line of sight by dust grains, but originally came from the HII
regions. But given the [SII] data that seems unlikely, although it may
be worthwhile to model this more accurately to place some limits on a
possible scattered light contribution. The [SII] data also make it
unlikely that the DIG is a collection of faint unresolved HII regions.
The HII region luminosity function of Kennicutt \etal (1989) suggests
that no more than 5\% of the DIG emission in the Sculptor galaxies
studied here can arise in unresolved HII regions.

The constancy of the diffuse fraction is surprising because of the
differences that exist in Hubble type and star formation activity of
the galaxies studied; quiescent spirals (M81 and M31), a vigorously
star forming Sbc (M51), a late-type Sd (NGC 300), and a starburst (NGC
253). An understanding of this situation cannot be reached without
better knowledge of the source of ionization of the DIG. The high
fractional H$\alpha$ luminosity for the DIG almost certainly implies
that {\it photoionization} has to dominate, but which stellar types
dominate?  The close association between DIG and HII regions makes it
tempting to think in terms of Lyc photons leaking from HII
regions. However, there is not yet independent evidence to support
this, and the lack of Lyc photons leaking from relatively unobscured
starburst nuclei (Leitherer \etal 1995) would perhaps indicate that
such leakage may not be very likely. The close association between DIG
and HII regions would, for example, also be expected if the DIG is
mostly ionized by field OB stars which are no longer associated with
discrete HII regions. Whether such stars can produce the required Lyc
photon rates needs to be determined from a careful census of such
stars in specific regions; we are not aware that any such studies have
been done for the galaxies in which the DIG has been analyzed. A
recent attempt was made by Patel \& Wilson (1995a \& b), however,
without having accurate data on the spectral types of the stars. We
are now working on several approaches. Nevertheless, in any case the
constancy of DIG fractional H$\alpha$ luminosities appears remarkable.

\acknowledgments 
We appreciate the help of the CTIO staff, in particular
Bill Weller, during the CTIO observing run. We acknowledge useful
discussions with Rob Kennicutt and Vanessa Galarza. This research was
supported by the NSF through grant AST-9123777, and by a Cottrell
Scholarship Award from Research Corporation. BEG was supported by a
grant from the New Mexico Space Grant Consortium.

\newpage

\figcaption{The continuum subtracted H$\alpha$ image of NGC 55. This
and the following images have been multiplied by $cos$ $i$ to put them
on a uniform brightness scale. The inclination of NGC 55 is
90$^\circ$, however, so the same correction used for NGC 253
($i$=78.5$^\circ$) was used here. The image is scaled logarithmically
to enhance the faintest emission levels. The greyscale ranges from
$-2$ to 500 pc cm$^{-6}$. We added a constant to the image so that the
entire range was positive, to enable the logarithmic scaling. The bar
at the lower left corresponds to 1 kpc at the distance of NGC
55. North is 18$^\circ$ clockwise from the top and East is 90$^\circ$
counter-clockwise from North. The diagonal pattern is a result of bias
structure on the chip (see \S 2.2).
\label{fig1}}

\figcaption{The continuum subtracted H$\alpha$ image of NGC 253,
scaled to face-on by multiplying the intensities with $cos$ $i$. The
circle near the southwest side of the galaxy corresponds to a poorly
subtracted bright star that was edited out. The levels vary from $-2$
to 500 pc cm$^{-6}$. The bar at the lower left corresponds to 1 kpc at
the distance of NGC 253. The image has been rotated so that North is
52$^\circ$ counter-clockwise from the top, and East is 90$^\circ$
counter-clockwise from North.  \label{fig2}}

\figcaption{The continuum subtracted H$\alpha$ image of NGC 300,
scaled to face-on by multiplying the intensities with $cos$ $i$. The
levels vary from $-6$ to 500 pc cm$^{-6}$. The lower value was chosen
to keep the noise at the same grey levels as for the inclined
galaxies, which were scaled differently due to the $cos$ $i$
correction. The bar at the lower left corresponds to 1 kpc at the
distance of NGC 300. North is up and East is on the
left. \label{fig3}}

\figcaption{(a) A masked image of NGC 253, using a cut at a constant
isophote of 80 pc cm$^{-6}$. The black areas on the galaxy are regions
that the masking technique defines as HII regions, where the pixel
value has been replaced by zero. This mask completely removes the
inner disk, even though significant diffuse emission is seen there in
figure 2. (b) This masked image was made using the median-subtracted
image method (see text \S 4.2). This mask is much better at
differentiating between the HII regions and DIG on the basis of
surface brightness, because the variation of the background DIG has
been removed in the masking process.  \label{fig4}}

\figcaption{Comparison of the effect of different isophotal cut-off
levels used in masking HII regions on the DIG fractional H$\alpha
+$[NII] luminosity. These curves correspond to the case where no
median filtering was used (c.f. figure 4a). The dashed lines
show the variation with continuum subtraction. At lower cut-off
levels, the diffuse fractions are comparable to those measured for the
Milky Way. Thus the relative power requirements of the DIG may well be
similar for the Sculptor galaxies and the Milky Way.  \label{fig5}}

\figcaption{Radial profiles of H$\alpha$ +[NII] emission for NGC 253
and NGC 300. The surface brightness is given in emission measure. Note
the difference in scale on the y-axis. Error bars are estimated from
the background uncertainty due to flat fielding errors.
Averaging azimuthally should reduce the uncertainty by at least a
factor of two, so the error bars here indicate half those
that would result from the estimated flat field uncertainty.
\label{fig6}}

\figcaption{The [SII]/(H$\alpha$ +[NII]) ratio for selected regions in
the Sculptor galaxies. Each point is the ratio in a 9''$\times$9''
box. In (a)-(c) (NGC 55, NGC 253, and NGC 300, respectively) each box
contains (part of) an HII region, and in (d)-(f) the boxes are placed
in the DIG. The solid line is the mean of all the points in bins of
200 pc cm$^{-6}$ for HII regions and 10 pc cm$^{-6}$ for
DIG. Representative error bars on the mean are shown in each
plot. They are a combination of rms error measured in the background
plus uncertainty due to the continuum subtraction. These are observed
values and are inclination dependent, which has no effect on the ratio
but does mean that the lower range of H$\alpha$ +[NII] emission
measure for NGC 300 (up to 40 pc cm$^{-6}$) should be compared with the
entire range (up to 150 pc cm$^{-6}$) for the inclined galaxies. This is
why there are no points at higher emission levels in the DIG in NGC
300.  \label{fig7}}

\end{document}